\documentclass[%
 aip,
 amsmath,amssymb,
 reprint,%
]{revtex4-1}

\usepackage{graphicx}
\usepackage{dcolumn}
\usepackage{bm}

\usepackage[utf8]{inputenc}
\usepackage[T1]{fontenc}
\usepackage{mathptmx}
\usepackage{etoolbox}
\usepackage{xcolor}

\newcommand{\met}{\mathrm}
\newcommand{\mat}{\mathbf}

\usepackage{array}
\usepackage{braket}

\makeatletter
\def\@email#1#2{%
 \endgroup
 \patchcmd{\titleblock@produce}
  {\frontmatter@RRAPformat}
  {\frontmatter@RRAPformat{\produce@RRAP{*#1\href{mailto:#2}{#2}}}\frontmatter@RRAPformat}
  {}{}
}%
\makeatother
\begin{document}

\preprint{AIP/123-QED}

\title[Photofragmentation of cyclobutanone at 200 nm: TD-DFT vs CASSCF electron diffraction]{Photofragmentation of cyclobutanone at 200 nm: TD-DFT vs CASSCF electron diffraction }
\author{Alberto Martín Santa Daría}
 \affiliation{Departamento de Qu\'imica F\'isica, Universidad de Salamanca https://ror.org/02f40zc51, 37008, Spain
 }
\author{Javier Hernández-Rodríguez}%
 
\affiliation{Departamento de Qu\'imica F\'isica, Universidad de Salamanca https://ror.org/02f40zc51, 37008, Spain
}%
\author{ Lea M. Ibele}%
\affiliation{ 
Université Paris-Saclay, CNRS, Institut de Chimie Physique UMR8000, 91405 Orsay, France
}%
\author{Sandra G\'omez$^*$}
 \email{sandra.gomez@usal.es}
\affiliation{Departamento de Qu\'imica F\'isica, Universidad de Salamanca https://ror.org/02f40zc51, 37008, Spain
}%

\date{\today}

\begin{abstract}
To simulate a 200 nm photoexcitation in cyclobutanone to the n-3s Rydberg state, classical trajectories were excited from a Wigner distribution to the singlet state manifold based on excitation energies and oscillator strenghts. Twelve singlet and twelve triplet states are treated using TD-B3LYP-D3/6-31+G$^{**}$ for the electronic structure and the nuclei are propagated with the Tully Surface Hopping method. Using TD-DFT, we are able to predict the bond cleavage that takes place on the S$_1$ surface as well as the ultrafast deactivation from the Rydberg n-3s state to the n$\pi^*$. After showing that triplet states and higher-lying singlet states do not play any crucial role during the early dynamics (i.e., the first 300 fs), the SA(6)-CASSCF(8,11)/aug-cc-pVDZ method is used as an electronic structure and the outcome of the non-adiabatic dynamic simulations is recomputed. Gas-phase ultrafast electron diffraction (GUED) spectra are computed for both electronic structure methods, showing significantly different results.  
\end{abstract}

\maketitle

\section{\label{sec:level1}Introduction}
\noindent
The exploration of photochemical molecular dynamics in theoretical chemistry has been a longstanding challenge, necessitating the simultaneous consideration of quantum mechanical effects on both nuclei and electrons. 
One of the main difficulties lies in predicting which degrees of freedom are activated upon radiation and this challenge falls within the dynamophore category, where the ultimate goal is to predict the nuclear motions and associated changes in electronic state and bonding environment.\cite{domcke2012role,evans2013applied,mai2020molecular,nelson2020non}

The photochemistry of cyclobutanone and its derivatives has been the subject of numerous studies, aiming to understand their physicochemical properties and their role in various reactions.\cite{morton1970molecular,umbricht1998photochemical,hue2003synthesis,karkas2016,zanini2021synthesis}
For instance, the photochemical rearrangement of cyclobutanone in [2 + 2] photocycloaddition plays a key role in the synthesis of advanced structures with relevance to biological systems.\cite{karkas2016}
Despite the wealth of knowledge on cyclobutanone photochemistry, predicting the outcomes of photoinduced processes in organic molecules remains a major challenge \cite{Solling2021}, due to its substantially different photochemical behavior compared to other cyclic and non-cyclic ketones.\cite{turro1972molecular_behaviour,lee1977laser_behaviour}

Cyclobutanone has been an attractive system since Benson and Kistiakowsky demonstrated in 1942 that its photolysis yields ethylene and ketene (C2 channel) in a 60\% yield, while the formation of C$_3$H$_6$ + CO (C3 channel) represents a 40\% yield.\cite{benson1942photochemical} Subsequent efforts have been made to understand the product formation after cyclobutanone is irradiated with light at experimental and theoretical levels.\cite{electronic_whitlock_1971,fluorescence_lee_1971,fluorescence_hemminger_1972,ring_drurylessard_1978,s1n_baba_1984,o1991vacuum,diau2001femtochemistry,symmetry_kuhlman_2012,Xia2015,Liu2016,coherent_larsen_2017,Kao2020}

In 2012, Kuhlman et al. studied the internal conversion from the 3s Rydberg state (the S$_2$ state at the Franck-Condon region) to the n$\pi^*$ (S$_1$) in cyclic ketones both theoretically and experimentally. \cite{Kuhlman2012-exp,symmetry_kuhlman_2012} Time-resolved mass spectrometry concluded that, for cyclobutanone, two time constants could be extracted, one at 80 fs and another one at 740 fs. The internal conversion among the states seemed to be lead by the ring-puckering vibrational mode. From the theoretical side, they constructed a linear vibronic coupling hamiltonian using five degrees of freedom and run quantum dynamics on the resulting potentials. The time constant for the internal conversion was found to be 950 fs. 

More recently, Diau et al. employed femtosecond time-resolved spectroscopy emphasizing the role of rapid motions, such as ring-puckering and CO out-of-plane wagging in the $\alpha$-cleavage dynamics taking place on the S$_1$ surface. Theoretical methods, including CASSCF, elucidated the $\alpha$-CC bond-dissociation pathway, concluding that triplets dominate at longer wavelengths. \cite{diau2001femtochemistry}
This study was extended to a solvent medium, \cite{Kao2020} focusing on Norrish Type-I $\alpha$-cleavage in the S$_1$ (n$\pi^*$ state), revealing insights into the competition between direct $\alpha$-cleavage and S$_1$-state relaxation, obtaining a timescale of 650 fs and the formation of ketene-containing compounds.
The C–C cleavage on the S$_1$ state and facile $\alpha$-cleavage in the triplet state after intersystem crossing from S$_1$ has been investigated via static computational methods exploring the S$_1$ and T$_1$ states, \cite{Xia2015} reporting a small energy barrier in the S$_1$ state.

Ab-Initio Multiple Spawning non-adiabatic dynamics simulations conducted by Liu et al. \cite{Liu2016} focused on the n$\pi^*$ deactivation and estimated the S$_1$ lifetime to be 484.0 fs. Their focus on singlet dynamics revealed a time constant for S$_1$ $\alpha$-cleavage of 176.6 fs, indicating the prevention of a statistical distribution of excess energies in the S$_1$ state. Experimental observations highlighted the decomposition of photo-excited cyclobutanone into C2 and C3 channels in the S$_0$ state, with a branching ratio dependent on the excitation wavelength.

In light of these considerations, the objective of this study is to further our understanding of cyclobutanone photochemistry through a detailed simulation of photochemical molecular dynamics. This has been motivated by an experiment at the SLAC Megaelectronvolt Ultrafast Electron Diffraction facility, where a gas sample of cyclobutanone is irradiated with 200 nm light and electron diffraction images are obtained. Electron diffraction is a powerful technique discovered in the early 20th century that allows researchers to follow the nuclear motion of molecules, providing valuable insights into ultrafast processes and structural dynamics at the atomic and molecular levels.\cite{Centurion2022} While waiting for the outcome from this experiment which will contribute to the existing body of knowledge on cyclobutanone photochemistry, the computational chemists, in particular the non-adiabatics community, have been challenged to predict it. The goal is to advance our understanding of excited state simulations and their predictive capabilities, potentially revolutionizing the design of light-driven molecular systems for applications ranging from renewable solar energy to bioimaging. \cite{blankenship2011comparing,staniforth2017first,simone2020organic} \\

The manuscript is organised as follows: after a short description on the theoretical methodology in Sec.~\ref{sec:methods}, the numerical results obtained for cyclobutanone as case of study are presented and discussed in Sec.~\ref{sec:results}. There we assess firstly the electronic structure methodology, and then we report the numerical results of the non-adiabatic dynamic calculations, including the analysis of the differences found for the different electronic structure methods. Finally, we present the experimental observables computed from the results of the non-adiabatic dynamics.
The general conclusions extracted from this work are summarized in Sec.~\ref{sec:conclusions}.

\section{\label{sec:methods}Methods and theoretical description}
\subsection{\label{sec:electronic}Electronic structure}
To properly account for states of different and mixed character (as it is the case in cyclobutanone with a manifold formed of valence and Rydberg states) one should in principle use multireference methods, preferably accounting for both static and dynamical correlation. The CASPT2 method, has been chosen over the years as the golden standard for electronic structure calculations. However, for studying the time evolution of systems after light irradiation, i.e., performing excited state dynamics in a manifold of coupled electronic states including non-adiabatic effects, it becomes computationally very expensive. 

A solution to this problem could be parametrising excited state potentials, such as those based on a linear vibronic coupling hamiltonian \cite{koe84:59,wor04:127} and run molecular dynamics on the coupled manifold of states. For bond cleavages and long range vibrational motions, this is nonetheless non-advisable and using vibronic coupling models may lead to unphysical dynamics on the precomputed potentials. \cite{Zobel2021,Penfold2023}

To run molecular dynamics on-the-fly, i.e., calculating energies, gradients and couplings while the molecule moves, the TD-DFT method is often an excellent choice since in many cases it offers a good compromise between accuracy and computational affordability. With the awareness that the DFT results are always influenced by the choice of the density functional parameters, there is a need of performing a careful benchmark against better electronic structure methods and experimental observables. 

In the case of cyclobutanone, the TD-B3LYP-D3/6-31+G$^{**}$ \cite{clark1983a,hariharan1973a,hehre1972a} method was proven to provide an excellent agreement with respect to the available experimental results while keeping the computational time feasible to run dynamic simulations. In the Table S3 in the ESI$^\dagger$, a full comparison against other methods, basis sets and functionals is made. The choice of the basis set is not surprisingly crucial to be able to describe Rydberg states correctly in this molecule, with other basis sets providing energies 2 eV apart from the correct result and not finding the n-3p Rydberg states. It should, however, be noted that for an accurate description of Rydberg states generally long-range corrected functionals are the best choice.\cite{tawada2004long,maitra2016perspective,tozer2000determination} However, benchmarks have shown that in specific cases the B3LYP functional was able to reproduce the energies of Rydberg states with reasonable accuracy. 
The Tamm-Dancoff approximation (TDA) \cite{hirata1999time,seidu2015applications} has been used for the TD-DFT calculations, using the ORCA 5.0 electronic structure program.\cite{Orca12,Orca22}

In linear response TD-DFT using the adiabatic approximation, the calculation suffers from instabilities in the vicinity of conical intersections between the ground and excited state due to the null coupling element between them\cite{todd, jack}, although it has been shown that in some instances the use of TDA helps to reduce these instabilities\cite{casida}.
Thanks to its computational affordability, we still decided to use TDA-TD-DFT to study the role of Rydberg states and triplets in the initial dynamics and the photofragmentation pathways that take place on the excited state. 

As more reliable methods, CASSCF and CASPT2 have been employed to have a better description of the excited state fragmentation of cyclobutanone. 
The openMOLCAS computer program \cite{openmolcas2020,openmolcas2023} has been used to calculate complete active space self-consistent field (CASSCF) \cite{ROOS1980157} and multistate complete active space second-order perturbation (CASPT2) \cite{FINLEY1998299} vertical energies. 
For both methods, the aug-cc-pVDZ basis was used. \cite{dunning1989a,kendall1992a}
The state-average CASSCF computations were performed using an active space of 8 electrons in 11 orbitals, CAS(8,11). 
This active space was constructed from an initial guess on the optimized structure including four electrons in two occupied orbitals, $n$ and the CO $\pi$, and five more virtual orbitals, the CO $\pi^*$ and four Rydberg (3s, 3p$_x$, 3p$_y$, and 3p$_z$). 
To this initial active space, we added, from  the conical intersection between S$_0$ and S$_1$ (Tab.~S1 of the ESI$^\dagger$), 
two $\sigma$ occupied orbitals and their corresponding two $\sigma^*$ virtual orbitals, defining our final CAS(8,11) active space (Fig.~S1 of the ESI$^\dagger$). 

We are including the first 6 electronic states in the state average CASSCF, SA(6)-CASSCF(8,11) and in the multistate MS(6)-CASPT2 calculations.
For the case of MS(6)-CASPT2, a 0.1 imaginary shift to the zero order Hamiltonian \cite{forsberg1997multiconfiguration} is employed in order to correct a low CASSCF reference weight in one of the higher lying states.

\subsection{Trajectory Surface Hopping\label{sec:tsh}}
The Tully surface hopping (TSH)\cite{Tully1971, Tully1990} mixed quantum-classical non-adiabatic dynamics method was our choice to propagate the nuclei, due to its ability to offer a good balance between computational effort and accuracy.

This method propagates classically the nuclei following the Newton equations of motion
\begin{equation}
  M_A\frac{\met{d}^2}{\met{d}t^2}\mat{R}_A(t)=-\frac{\met{d} V_{\alpha}(\mat{R})}{\met{d} \mat{R_A}} ,
  \label{eq:theo:newton}
\end{equation}
where the atom $A$ moves, changing its position with time ($\mat{R}(t)$), on single potential energy surfaces ($V_{\alpha}\left(\mat{R}(t)\right)$) driven by the surface gradients. Compared to more exact nuclear wavefunction based methods, 
TSH consists of individual trajectories that move completely independent from each other. 
As a result of the independent trajectory approximation, even in the limit of infinite trajectories, TSH is not guaranteed that the dynamics converge to follow an exact wavepacket evolution, although it has been shown in many cases to be a close approximation.

The velocity Verlet algorithm\cite{Verlet1967} propagates the system from one-time step to another, and velocities $\mat{v}_A(t)$ and atomic coordinates $\mat{R}_A(t)$ evolve as
\begin{align}
  \mat{R}_A(t+\Delta t) &= \mat{R}_A(t) + \mat{v}_A(t)\Delta t + \frac{1}{2M_A} \frac{\met{d} V_{\alpha}(\mat{R})}{\met{d} \mat{R_A}} \bigg|_{\mathbf{R_A}=\mathbf{R}_A(t)}  \Delta t^2, \\ 
  \mat{v}_A(t+\Delta t) &= \mat{v}_A(t) + \frac{1}{2M_A}\frac{\met{d} V_{\alpha}(\mat{R})}{\met{d} \mat{R_A}} \bigg|_{\mathbf{R_A}=\mathbf{R}_A(t)}\Delta t \nonumber \\ 
  & +\frac{1}{2M_A}\frac{\met{d} V_{\alpha}(\mat{R})}{\met{d} \mat{R_A}}\bigg|_{\mathbf{R_A}=\mathbf{R}_A(t+\Delta t)}\Delta t.
\end{align}

The nuclei are assumed to move on the same electronic surface -adiabatically- most of the time and the non-adiabatic coupling among states is monitored. If a large coupling with other electronic state is detected, the trajectory undergoes a non-adiabatic transition known as hop, thus the name surface hopping.\cite{Tully1971, Tully1990} At every time step, the trajectories have the choice to change their electronic state based on a probability to hop that depends on the coefficients of the electronic states involved. The electronic wavefunction is expressed in a basis of the adiabatic electronic states with time-dependent expansion coefficients. When this wavefunction ansatz is inserted into the time-dependent Schr\"odinger equation, it yields to the following equations of propagation for the coefficients:

\begin{align}
  & \frac{\met{d}c_{\beta}(t)}{\met{d}t}  = \nonumber \\
  & -\sum_{\alpha}\Bigg[\met{i}\underbrace{\Braket{\Psi_{\beta}^{\mathrm{el}}|\hat{H}^{\met{el}}|\Psi_{\alpha}^{\mathrm{el}}}}_{ H_{\beta\alpha}}+\underbrace{\Braket{\Psi_{\beta}^{\mathrm{el}}|\frac{\met{d}}{\met{d}t}|\Psi_{\alpha}^{\mathrm{el}}}}_{ K_{{\beta}\alpha}} \Bigg]c_{\alpha}(t) \; . \label{eq:theo:coeffs}
\end{align}

The Hamiltonian term is directly calculated from the analytical form of the potential energy surface, if a precomputed surface is used, or on-the-fly via electronic structure methods. The second term is calculated as the non-adiabatic coupling between electronic states and the atomic velocities: $K_{\beta\alpha}= \Braket{\Psi_{\beta}^{\mathrm{el}}|\nabla_R|\Psi_{\alpha}^{\mathrm{el}}}\mat{v}_R$.
In this work we used the SHARC implementation of TSH\cite{Mai2015b} from the Gonz\'alez research group. 
The TSH calculations used the ``fewest switches'' algorithm,\cite{Tully1990} along with the energy-based decoherence correction.\cite{granucci2007critical,Granucci2010} The nuclear time-step was chosen to be 0.5 fs; 25 substeps were used for the electronic integration. 
The electronic wavefunction is propagated using the local diabatization algorithm and consequently, the hopping probabilities were obtained from wavefunction overlaps.\cite{granucci_direct_2001,Plasser2016} 

500 initial structures and velocities were generated from a Wigner distribution and vertical excitation energies and oscillator strengths were computed to simulate the absorption spectrum.
Starting from these initial conditions, trajectories were selected based on the oscillator strengths whenever the excited state energies fall into the excitation window around the absorption spectra maximum which corresponds to the n-3s transition (approximately 200 nm, the experimental excitation wavelength). 

For the assessment of the role of triplet states during the dynamics, we chose the TD-B3LYP-D3/6-31+G$^{**}$ electronic structure method, including 12 coupled singlet electronic states (S$_0$--S$_{11}$) and 12 triplet electronic states (T$_1$--S$_{12}$). The spin-orbit coupling is calculated using the ZORA Hamiltonian.\cite{lenthe1993relativistic,lenthe1994relativistic,lenthe1996zero}  This large number of electronic states (48 in total, since SHARC works in the full adiabatic picture - eigenstates of the electronic hamiltonian including spin-orbit coupling terms) was selected since they fall in the same energy range, as observed in the density of states plot in Fig.~S3 of the ESI$^\dagger$. 
These results, discussed in Sec.~\ref{sec:dyn}, confirm that triplet electronic states do not play a role during the dynamics, i.e., their population is negligible.  This allows us to perform TSH simulations with the SA(6)-CASSCF(8,11) method including only the energetically relevant singlet states. 

As mentioned before, the coupling with the ground state is incorrectly described with TD-DFT, so we re-run TD-B3LYP-D3/6-31+G$^{**}$ trajectories, although this time including only six singlet states and no triplet states, forcing the system to decay to the ground state when a difference of 0.1~eV was found between S$_1$ and S$_0$.

\subsection{Electronic scattering\label{sec:scattering}}
The total scattering intensity, $I(s)$, can be decomposed into two contributions, atomic and molecular scattering. (Eq.\ref{eq:scattering_int}) 

\begin{equation}
  I(s) = I_{at}(s) +I_{mol}(s)
  \label{eq:scattering_int}
\end{equation}

The atomic scattering term, $I_{at}(s)$, is defined as the sum of each atomic differential cross-section and it does not contain structure information: 

\begin{equation}
  I_{at}(s) = \sum^N_{i=1}|f_i(s)|^2
  \label{eq:scatt_int_at}
\end{equation}

The scattering amplitudes are calculated using ELSEPA software.\cite{salvat2005elsepa} The molecular scattering contribution, is expressed as the sum of interference terms of each possible atom pair. For that, the internuclear distances in the target molecules are necessary. 

\begin{equation}
  I_{mol}(s) = \sum^N_{i=1} \sum^N_{j \neq 1} |f_i(s)||f_j(s)|\frac{sin(sr_{ij})}{sr_{ij}}
  \label{eqscatt_int_mol}
\end{equation}

where $f_i$ and $f_j$ are the elastic scattering amplitudes for the $i$ and $j$ atoms and $r_{ij}$ is the internuclear distance between the $i$ and $j$ atoms. 

We present the simulated UED signals in terms of modified scattering intensity which increase the oscillations in the $I_{mol}$ contribution and suppress the rapid decrease in scattering intensity arising from the $s^{-2}$ scaling of $f_i(s)$.
The modified scattering intensity is defined as: 
\begin{equation}
  sM(s) = \frac{I_{mol}(s)}{I_{at}(s)}s
  \label{eq:mod_scatt_int}
\end{equation}
The $sM(s)$ curve is descomposed into a pair-distribution function (PDF) of all contributing interatomic distances, 

\begin{equation}
  PDF(r) = \int^{S_max}_{0} sM(s) sin(sr) e^{-ks^2} ds
  \label{eq:pdf}
\end{equation}

where $s_{max}$ is the maximum transfer in the diffraction with adequate signal-to-noise ratio, $r$ is the internuclear distance between atom pairs, and $k=0.03$ is a damping factor used to suppress the high $s$ contribution smoothly to zero.\cite{figueira2023monitoring} 

The time-dependent signals are reported as $\Delta PDF(r,t)$ which is calculated from the $\Delta sM(s,t) = sM(s,t) - sM(s,t=0)$ as:
\begin{equation}
\Delta  PDF(r,t) = \int^{S_max}_{0} \Delta sM(s,t) sin(sr) e^{-ks^2} ds
\end{equation}
where the signal of $sM(s,t=0)$ is calculated from the average of all trajectories at time $t=0$.

\section{Results and discussions\label{sec:results}}
\subsection{\label{sec:res_vertical}Validation of the electronic structure}
The first reported spectrum of cyclobutanone dates from 1973, when Hemminger and coworkers \cite{hemminger1973laser}
recorded the gas-phase absorption spectrum on the $n \rightarrow \pi^*$ region involving the transition to the first singlet excited state, S$_1$. 
In 1991, O'Toole \emph{et. al.} \cite{o1991vacuum} reported the vacuum-ultraviolet spectrum characterizing the transition involving the Rydberg states of this molecule. 
These two experimental spectra, and their review in Ref. \citenum{diau2001femtochemistry}, has been used as reference to choose and validate the electronic structure methodology used throughout this work.  

\begin{figure}
  \begin{center}
    \includegraphics[width=0.90\linewidth]{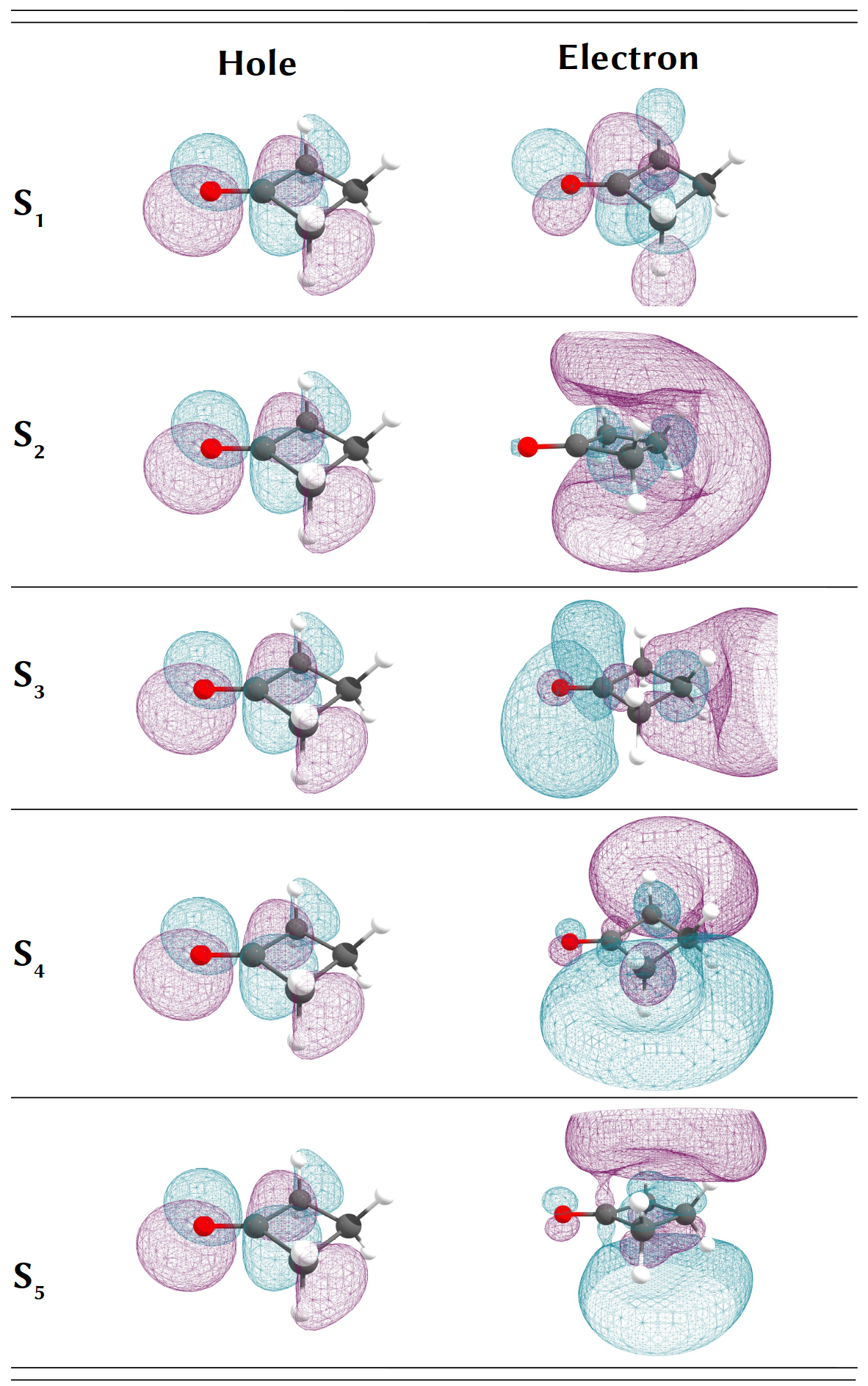}
    \end{center}     
    \caption{%
      Natural transition orbitals (NTOs) for the first five excited singlet states (S$_1$--S$_5$) of cyclobutanone calculated with TD-B3LYP-D3/6-31+G$^{**}$.
      For each state, the hole orbital is on the left and the particle (electron) orbital is on the right. The NTOs were recalculated with SA(6)-CASSCF(8,11)/aug-cc-pVDZ and were found to have the same electronic character. 
      }
    \label{fig:ntos}
\end{figure}

For the first 5 singlet electronic states, vertical energies with their corresponding oscillator strengths computed at the Frank-Condon region have been included in Table~\ref{tab:energies}, specifying the symmetry and the molecular orbitals involved (character) for every excited state computed. This table contains the results for the three electronic structure methodologies described in Sec.~\ref{sec:electronic}: TD-B3LYP-D3/6-31+G$^{**}$, SA(6)-CASSCF(8,11)/aug-cc-pVDZ and SA(6)-CASPT2(8,11)/aug-cc-pVDZ.
\begin{table*}
\centering
\caption{Vertical energies in eV and their corresponding oscillator strengths in brackets computed using three different levels of theory: TD-B3LYP-D3/6-31+G$^{**}$, 
SA(6)-CASSCF(8,11)/aug-cc-pVDZ and 
SA(6)-CASPT2(8,11)/aug-cc-pVDZ.
The references of the experimental works are provided in the corresponding entrance of the table.
\label{tab:energies}
}
\begin{tabular}{c c l c r l@{\ \ }  c r l@{\ \ }  c |  c c l c  r l@{\ \ } c }
\hline\hline\\[-0.2cm]
 State & Symmetry & Character & &  \multicolumn{2}{c}{TD-B3LYP-D3} & & \multicolumn{2}{c}{CASSCF} &&
 State & Symmetry & Character  && \multicolumn{2}{c}{CASPT2} & Experiment \\[0.1cm]
\hline\\[-0.1cm]
S$_1$ & 1A'' & n$\rightarrow\pi^*$              &&   4.25 & (0.000) & & 5.64 & (0.000)&&S$_1$ & 1A'' & n$\rightarrow\pi^*$             && 3.94 & (0.000) & 4.42 [Ref. \citenum{hemminger1973laser}] \\[0.1cm]
S$_2$ & 2A'' & n$\rightarrow3\text{s}$          &&   6.30 & (0.071) & & 5.68 & (0.056)&&S$_2$ & 2A'' & n$\rightarrow3\text{s}$         && 6.11 & (0.043) &  6.29 [Ref. \citenum{o1991vacuum}] \\[0.1cm]
S$_3$ & 3A'' & n$\rightarrow3\text{p}_\text{z}$ &&   6.67 & (0.001) & & 6.40 & (0.003)&&S$_3$ & 3A'' & n$\rightarrow3\text{p}_\text{y}$ && 6.72 & (0.000) & 6.94 [Ref. \citenum{o1991vacuum}]\\[0.1cm]
S$_4$ & 1A'  & n$\rightarrow3\text{p}_\text{x}$ &&   6.91 & (0.009) & & 6.48 & (0.009)&&S$_4$ & 4A'' & n$\rightarrow3\text{p}_\text{z}$ && 6.82 & (0.001) & 6.94 [Ref. \citenum{o1991vacuum}]\\[0.1cm]
S$_5$ & 4A'' & n$\rightarrow3\text{p}_\text{y}$ &&   7.07 & (0.000) & & 6.58 & (0.000)&&S$_5$ & 1A' & n$\rightarrow3\text{p}_\text{x}$ && 6.96 & (0.007) &  6.94 [Ref. \citenum{o1991vacuum}]\\[0.1cm]
\hline\hline
\end{tabular}
\end{table*}
At the same time, for these first five singlet electronic states, Fig.~\ref{fig:ntos} shows the representation of the corresponding occupied (hole) and virtual (electron) natural transition orbitals (NTOs) computed with TD-B3LYP-D3/6-31+G$^{**}$ , which are almost identical to the ones calculated using SA(6)-CASSCF(8,11)/aug-cc-pVDZ level of theory (see the comparison of the NTOs in Fig.~S2 of the ESI$^\dagger$). 
It is important to mention that the orbital ordering of the 3p Rydberg states is different in CASPT2, while the n-$\pi^*$ and n-3s transitions remain dominant in the first two singlet excited electronic states for the three electronic structure methods.

\begin{figure}
  \begin{center}
    \includegraphics[width=0.8\linewidth]{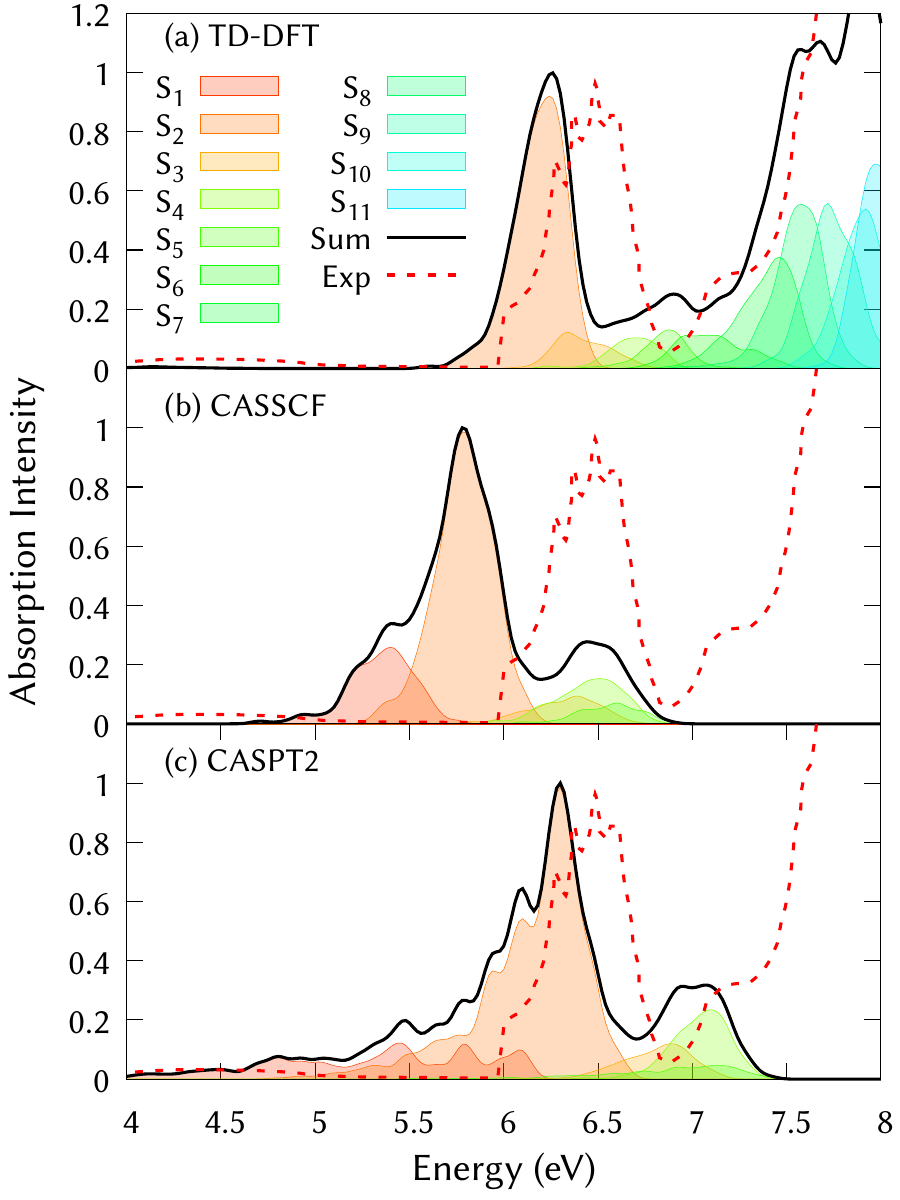}
    \end{center}     
    \caption{%
      Vertical absorption spectrum calculated with three levels of theory: 
      (a)~TD-B3LYP-D3/6-31+G$^{**}$, 
      (b)~SA(6)-CASSCF(8,11)/aug-cc-pVDZ, and
      (c)~SA(6)-CASPT2(8,11)/aug-cc-pVDZ.
      Exciting 500 initial conditions from a Wigner distribution for each method.
      The experimental absorption spectra, red dashed line, are taken from Refs.~\citenum{diau2001femtochemistry,o1991vacuum,hemminger1973laser}. The experimental intensities in the region from 4 to 6 eV have been multiplied by a factor 10 to boost their visibility.}
      
    \label{fig:abs_spec}
\end{figure}

Using the same three electronic structure approaches, we computed vertical absorption spectra using 500 initial conditions from a Wigner distribution.\cite{Crespo-Otero2012} The three resulting spectra have been represented in the panels of Fig.~\ref{fig:abs_spec}.
Although the three theoretical methodologies can reproduce every feature of the experimental spectra, based on these static results our method of choice would have been CASPT2, which shows a deviation with respect to experiment of $\sim$6~nm (0.2~eV) in the position of the brightest absorption peak (S$_2$).

Considering that TDDFT shows a good agreement (0.4~nm) with respect to the experimental (n,$\pi^*$) peak, this method has been chosen to run the on-the-fly non-adiabatic dynamics over CASPT2 due to its high computational cost. 
As commented before, in order to better describe the Rydberg states and the photo-fragmentation of the system, the same dynamics has been attempted using CASSCF as reference method where the ground state is coupled with the excited states. Comparing the dynamics with TDDFT and CASSCF will allow us to understand the effect that an accurate description of the Rydberg states will have on the consequent dynamics.

It is interesting to note that, while the n$\rightarrow\pi^*$ transition is symmetry forbidden and fully dark for all approaches at the Franck-Condon geometry, this symmetry is partially broken in the CASSCF and CASPT2 spectra, giving rise to a low-energy shoulder and tail feature, respectively.
Even though CASSCF shows the largest differences with respect to the experimental reference, 21~nm (0.7~eV) in the FC region, we tested this method with respect to CASPT2 along a reaction path from the optimized geometry of S$_0$ to the optimized conical intersection (CI) of S$_0$ and S$_1$ (see Figure S4 at the ESI$^\dagger$). 
With this analysis we concluded that with the CASSCF method we obtain surfaces that are parallel to those obtained with CASPT2. Therefore, CASSCF can be used instead to be able to run the dynamics in a reasonable period of time.

\subsection{Non-adiabatic dynamics from the n-3s Rydberg state\label{sec:dyn}}
After the assessment of the level of theory based on the static results, TD-B3LYP-D3/6-31+G$^{**}$ was chosen to run TSH non-adiabatic dynamics including the first 12 singlet and the first 12 triplet electronic states.
The trajectories were run until 250 fs. Despite the fact that TDDFT is a computationally affordable method, and due to the rather high density of states in this energy range, the trajectories were computationally demanding, taking each two days in 8 CPUs. 
The excitation window for these trajectories was chosen to be 5.9--6.5 eV, corresponding to exciting the cyclobutanone molecule to its Rydberg n-3s state, based on the absorption spectrum computed with TD-B3LYP-D3/6-31+G$^{**}$ in Fig.~\ref{fig:abs_spec}(a). 

\begin{figure}
  \begin{center}
    \includegraphics[width=\linewidth]{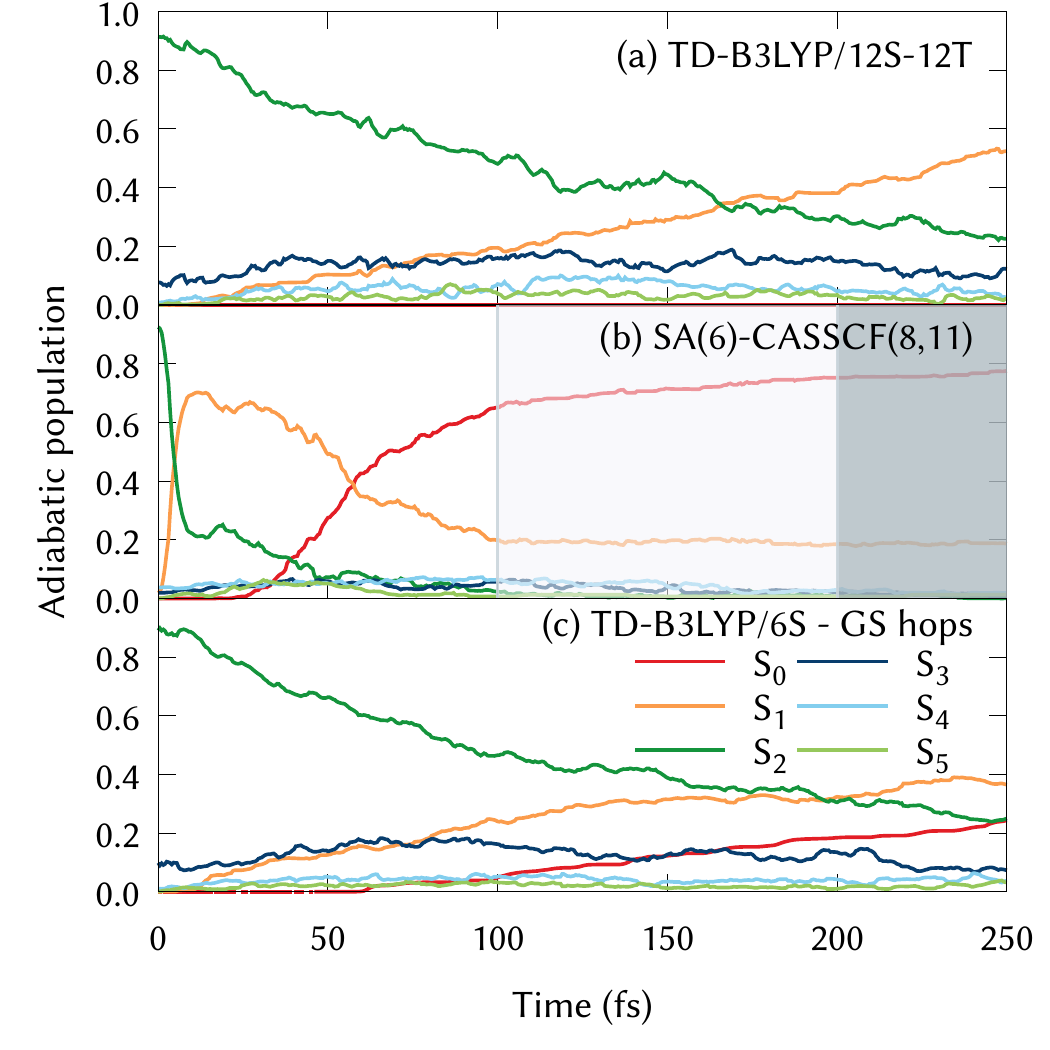}
    \end{center}     
    \caption{%
      Adiabatic population transfer after the excitation to the n$\rightarrow$3s (S$_2$) band of cyclobutanone using the on-the-fly Tully Surface Hopping method with three approaches:
       (a) TD-B3LYP-D3/6-31+G$^{**}$ with 118 trajectories, considering twelve singlet and twelve triplet states 
       (b) SA(6)-CASSCF(8,11)/aug-cc-pVDZ with 294 trajectories and 
       (c) TD-B3LYP-D3/6-31+G$^{**}$ on a six singlet state manifold and forcing the system to hop to the ground state when the difference between S$_0$ and S$_1$ is less than 0.1 eV, considering 289 trajectories. 
       }
    \label{fig:3pops}
\end{figure}

An analysis of the adiabatic population transfer was carried out with an ensemble of 118 trajectories, represented in panel (a) of Fig.~\ref{fig:3pops}.
The population is summed up for the triplet manifold, being its contribution less than 0.5$\%$ of total population. Based on this dynamics, we can conclude that Spin-Orbit couplings (SOCs) are negligible during this time scale and thus, for the other electronic structure methods, trajectories were run on the singlet manifold including only the first 6 singlet electronic states (S$_0$--S$_5$).

\begin{figure*}
  \begin{center}
    \includegraphics[width=\linewidth]{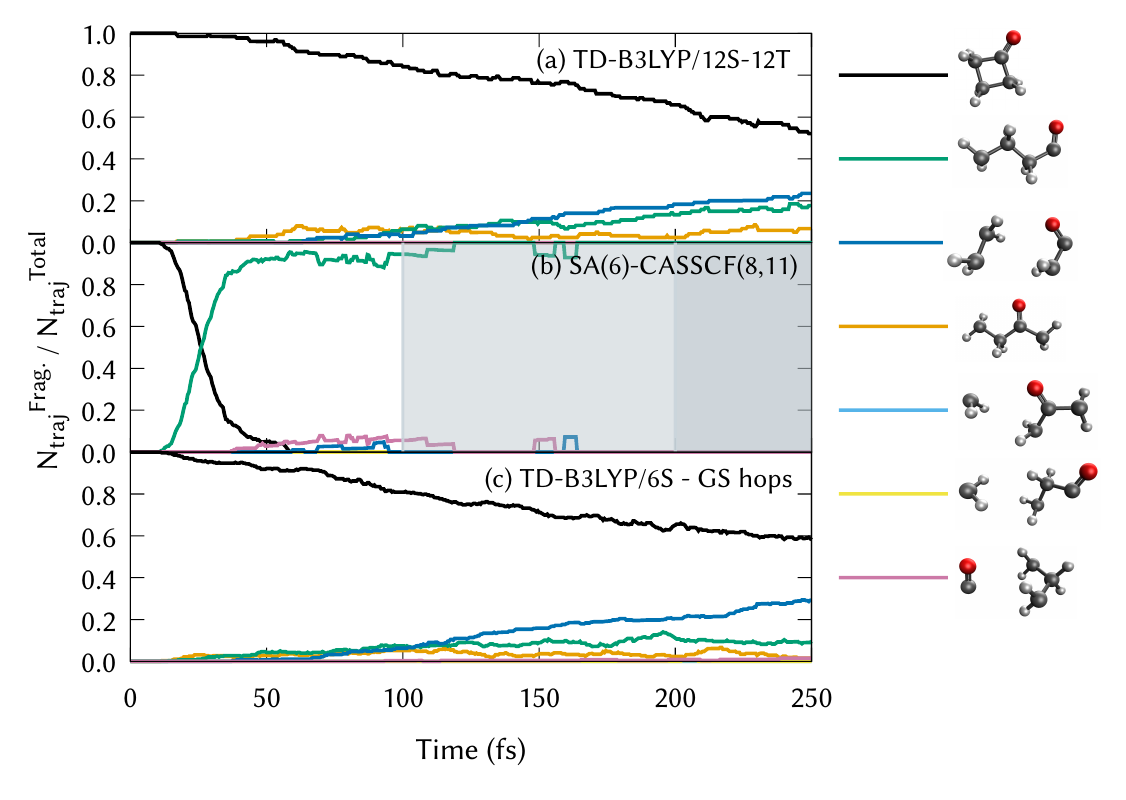}
    \end{center}     
    \caption{%
      Photodissociation products after the excitation of cyclobutanone to the energy window corresponding to excite to the n3s (S$_2$) band using 
      on-the-fly Tully Surface Hopping classical trajectories (TSH) with three different methods:
       (a) TD-B3LYP-D3/6-31+G$^{**}$, considering twelve singlet and twelve triplet states 
       (b) SA(6)-CASSCF(8,11)/aug-cc-pVDZ, and
       (c) TD-B3LYP-D3/6-31+G$^{**}$ considering six singlet states and forcing the system to hop to the ground state when the difference between S$_0$ and S$_1$ is less than 0.1 eV. 
    The number of trajectories is normalized at every time step. The shadow in grey shows when the number of trajectories fall under a threshold where we consider they may not be statistically converged. The first line (at 100 fs) marks a total of 30 trajectories and the second line (at 200 fs) marks 15 trajectories.  
      }
    \label{fig:photoproducts}
\end{figure*}

Problems in energy conservation are detected when CASSCF is used to evaluate the electronic energy. The orbitals included in the active space in CAS methods are chosen at the equilibrium geometry; these orbitals are optimized at each time step. Once the molecule is far distorted from its equilibrium geometry, the initial set of orbitals might not offer a sufficient description anymore leading to instabilities in the electronic structure.

During a TSH trajectory, the initial guess for the CASSCF orbitals is read from the previous time step calculation. If the structure between two consecutive steps is fairly different, as usual when a hop to another state occurs and the system follows a different state gradient, the orbitals may not be good enough for the new nuclear configuration and then an energy drift happens and the trajectory stops at this time step.
Due to this fact, the statistics of our CASSCF trajectories are far from the ones obtained with TDDFT. In TSH-CASSCF, out of a total of 294 trajectories initialized from the 500 initial conditions, 150 (51$\%$) survived up to 50 fs, 36 (12$\%$) up to 100 fs, 15 (5$\%$) up to 150 fs and 4 (1$\%$) up to 250 fs. 
However, we benchmarked the population transfer based on the number of trajectories included in the analysis ensemble (see Figure S7 in the ESI$^\dagger$) for the TSH-CASSCF results, obtaining a qualitatively correct behaviour for 50, 40 and 30 trajectories. 
The adiabatic population transfer obtained with TSH-CASSCF is represented in Fig.~\ref{fig:3pops}(b). Even when excluding the gray part of the figure where the number of trajectories is less than 30, the differences with respect to TDDFT are noticeable. 
However, one of the main differences between panels (a) and (b) in Fig.~\ref{fig:3pops} is that CASSCF allows the system to hop to the S$_0$, where we expect some of the dissociation processes to occur. 

To have a real comparison between the two methods, as mentioned above, we repeated the TSH-TDDFT simulations forcing the system to hop to the ground state when the difference between S$_1$ and S$_0$ was less than 0.1 eV. The corresponding adiabatic population transfer is included in Fig.~\ref{fig:3pops}(c). 

Comparing panels (b) and (c) in Fig.~\ref{fig:3pops} confirms the fact that non-adiabatic dynamics with TDDFT are 10 times slower than with CASSCF, which can be explained by the fact that the vertical energy difference between S$_1$ and S$_2$ states is smaller in CASSCF. 
This difference between both methodologies may indicate a limitation in the active space size in CASSCF, which may not be able to describe the  $\alpha$-cleavage of the ring.

To further analyze the dissociation of the cyclobutanone along the dynamics, we followed the geometries over time and classified every geometry in seven different photoproducts based on the interatomic distances between all carbon atoms in the ring. 
If the first bond breaking is an $\alpha$-CC, no aliphatic chain is generated and the product is a 4 carbon chain with the C=O terminal group, called here \textbf{C0$^{T}$} photoproduct. 
Instead, if we first have a $\beta$-bond cleavage, then the C=O is not terminal and we named it \textbf{C0}. 
When a second bond dissociation occurs, several pathways have to be considered. If the product is COC$_2$H$_4$ + CH$_2$, we again distinguish between the terminal C=O, \textbf{C1$^{T}$} and the acetone radical, \textbf{C1}.
One unique possibility appears for a 2 carbon aliphatic chain, COCH$_2$ + C$_2$H$_4$, named \textbf{C2} channel, following the nomenclature used in literature.\cite{benson1942photochemical,Liu2016}
Lastly, if both $\alpha$ bonds are broken, the product is CO + C$_3$H$_6$, called \textbf{C3} channel.
Every dissociation channel and the corresponding bond cleavage are schematized in Fig.~S5 of the ESI$^\dagger$.

Photodissociation channels of cyclobutanone calculated with the 3 methods are represented in Fig.~\ref{fig:photoproducts}, where the normalized number of trajectories ending up in a specific photoproduct is plotted for every time step during the TSH dynamic simulations.
Interestingly, panels Fig.~\ref{fig:photoproducts}(a) and Fig.~\ref{fig:photoproducts}(c) are very similar, both showing the photodissociation on TDDFT on-the-fly surfaces. The main difference, apart from the exclusion of the triplet manifold, is that the ground state is also populated after 60~fs in panel (c). 
This behavior is almost identical, although at 250~fs the population going through \textbf{C0$^{T}$} channel is half in panel Fig.~\ref{fig:photoproducts}(c) than in (a), which implies that going to the S$_0$ does not affect much the cleavage of the structure but, if the system is `forced' to stay on the S$_1$ there is more proportion of the \textbf{C0$^{T}$} and \textbf{C2} products. 
With TD-DFT the deactivation channels \textbf{C3}, \textbf{C1} and \textbf{C1$^{T}$} are negligible while the \textbf{C0} product appears with residual contribution.

Similarly to the analysis of the adiabatic populations, the dissociation pattern after excitation is quite different among CASSCF and TD-DFT, panels Fig.~\ref{fig:photoproducts}(b) and Fig.~\ref{fig:photoproducts}(c) respectively. 
In CASSCF, the $\alpha$-cleavage occurs much faster and before 50~fs almost all cyclobutanone have been broken to form the \textbf{C0$^{T}$} product while the percentage of the rest keeps below 10$\%$. 
After the $\alpha$-cleavage, we observed a second $\alpha$-CC bond breaking to form the \textbf{C3} product and the first $\beta$-cleavage to product \textbf{C2}. Again, both \textbf{C1} and \textbf{C1$^{T}$} products are not visible during this dynamics. 
Comparing Figures~\ref{fig:3pops} and ~\ref{fig:photoproducts}, we can try to relate the S$_2$ and \textbf{CB} populations. In TSH-CASSCF, the ring opening of \textbf{CB} starts taking place after 15 fs, right after the S$_2$ to S$_1$ internal conversion. Same conclusions can be extracted in TSH-TDDFT, where the \textbf{CB} decay is also slower than the S$_2$ internal conversion, indicating that the molecule needs to go from the Rydberg n-3s state to the n$\pi^*$ to begin the dissociation process. TD-B3LYP-D3 seems to agree better with previous experimental and theoretical findings, reporting that the lifetime of the Rydberg state is 740 fs and that the S$_1$ lifetime is about 500 fs.

\subsection{Experimental observables: Electron diffraction spectra }
\begin{figure}[h]
  \begin{center}
    \includegraphics[width=1\linewidth]{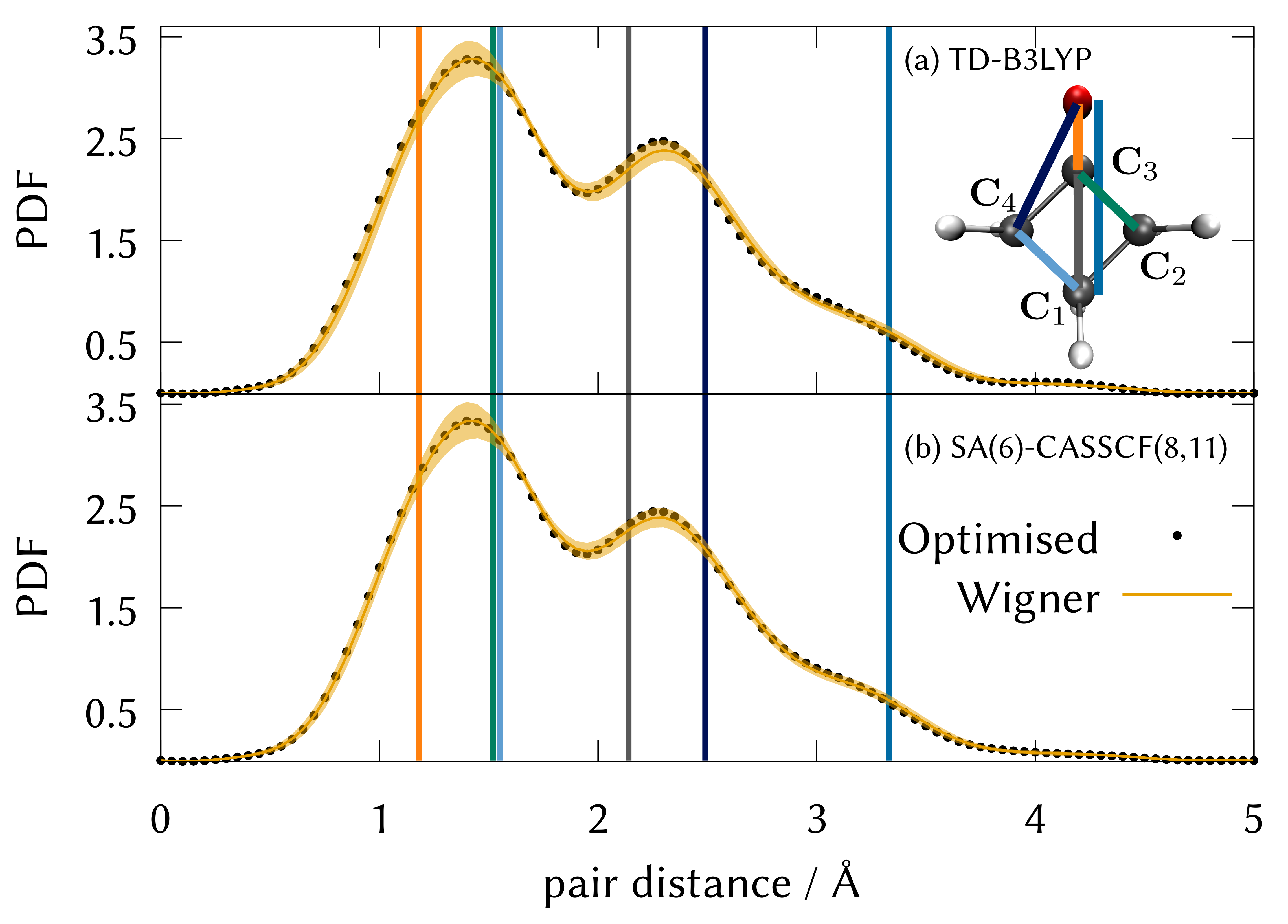}
    \end{center}     
    \caption{%
     Pair distribution function of cyclobutanone calculated with (a) TD-B3LYP-D3/6-31+G$^{**}$ and (b) SA(6)-CASSCF(8,11)/aug-cc-pVDZ. Black: optimised structure, orange: average structure from 500 wigner distribution structures. 
      }
    \label{fig:staticsig}
\end{figure}

Let us have a look at the electron diffraction quantities predicted from these dynamics. First, we calculated the steady-state atomic pair distribution functions (PDF) for both electronic structure methods, as shown in Fig.~\ref{fig:staticsig}. The PDFs are calculated from the optimized geometry as well as from the average ensemble of Wigner sampled geometries. Both methods yield only insignificant differences in the PDFs when compared with each other. The sampling effect does not affect the PDFs significantly. The PDF show three main features, the most intense peak centered at 1.4 \AA, a less intense peak at 2.3 \AA~and a shoulder-like feature at 3.2 \AA. Comparing with the atom distances at the optimized geometries, as indicated in Fig.~\ref{fig:staticsig}, the peak at 1.4 \AA~seems to arise from the C-O distance of the carbonyl group as well as the neighboring C-C distances of the ring. The second peak stems from the distance between the oxygen and the second carbons of the ring as well as the distance of the carbonyl carbon with the opposite carbon of the ring. Finally, the shoulder around 3.2 \AA~is related to the distance of the oxygen to the farthest carbon. 

\begin{figure}[h]
  \begin{center}
    \includegraphics[width=0.90\linewidth]{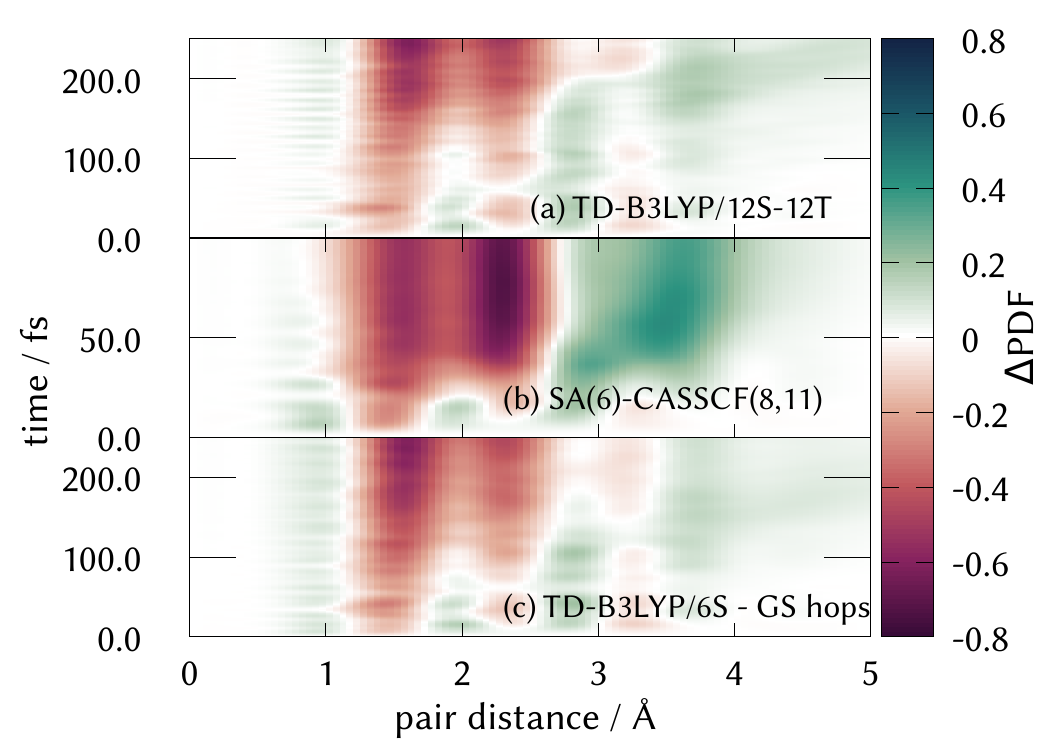}
    \end{center}     
    \caption{%
     Difference pair distribution functions of cyclobutanone calculated with (a) TD-B3LYP-D3/6-31+G$^{**}$ (b) SA(6)-CASSCF(8,11)/aug-cc-pVDZ (c) TD-B3LYP-D3/6-31+G$^{**}$ considering only three singlet states and forcing the system to hop to the ground state when the difference between S$_0$ and S$_1$ is less than 0.1 eV. For the trajectories that finished earlier, the last geometries have been continued until 250 fs. 
      }
    \label{fig:tdepsig}
\end{figure}

Subsquently, let us look at the time-dependent PDFs along the trajectories. Fig.~\ref{fig:tdepsig} shows the change of PDF over time. 
Both TDDFT signal traces show a weak depletion around 1.4 \AA~that starts almost instantly. After a small time delay, also the peak abound 2.3 \AA~is depleted. In addition, the signal at 3.2 \AA~is very weakly depleted initially but seems to oscillate.
The time trace from the CASSCF dynamics look very different. It is important to note that Fig.~\ref{fig:tdepsig} depicts the CASSCF traces only up to 100 fs, while the TDDFT traces are shown up to 250 fs. In the CASSCF dynamics, the depletion of the two main signals starts almost immediately after photoexcitation and the depletion increases significantly faster and stronger in comparison to the TDDFT traces which is in agreement with the much fast ring-opening observed. 
The spectrum shows an immediate depletion of the signal around 1.4 \AA, with around 10 fs delay the depletion of the signal at 2.3 \AA~starts. This indicates a ring-opening process where first the C$_3$-C$_2$ bond is broken by an initial motion of the C$_2$ carbon. Only with a small delay, the C$_3$=O carbonyl group moves away, which is shown by the delayed depletion around 2.3 \AA. 
A curious feature, which is not observed with TDDFT, is the increase of the signal between 3 and 4 \AA. Within the first 50 fs this signal increases around 3 \AA~subsequently it gets shifted to higher distances closer to 4 \AA.

\begin{figure}[h]
  \begin{center}
    \includegraphics[width=\linewidth]{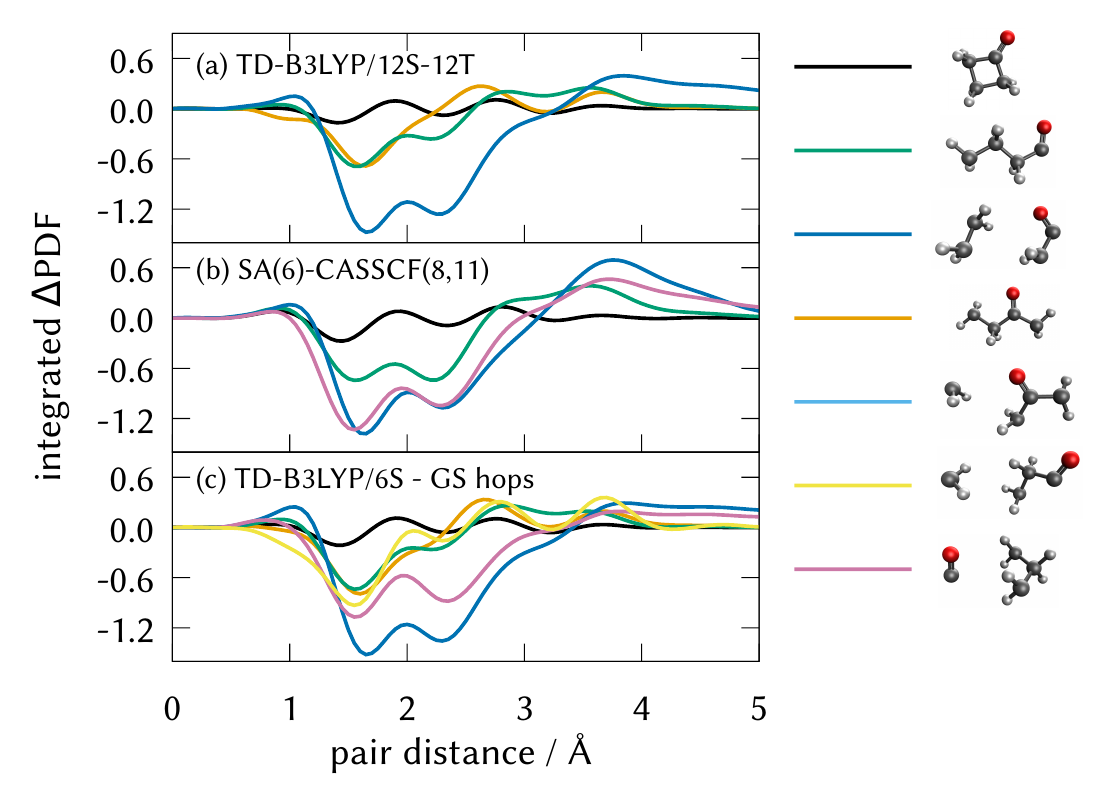}
    \end{center}     
    \caption{%
     Integrated difference pair distribution functions of different photoproducts of cyclobutanone calculated with (a) TD-B3LYP-D3/6-31+G$^{**}$ (b) SA(6)-CASSCF(8,11)/aug-cc-pVDZ (c) TD-B3LYP-D3/6-31+G$^{**}$ considering only three singlet states and forcing the system to hop to the ground state when the difference between S$_0$ and S$_1$ is less than 0.1 eV. 
      }
    \label{fig:photoprodsig}
\end{figure}

In Fig.~\ref{fig:photoprodsig}, we report the integrated $\Delta$PDF signal for all the photoproducts formed during the dynamics following the classification explained above.  All photoproducts show a depletion around 1.5 \AA. Interestingly, from the TDDFT dynamics we can calculate the expected $\Delta$PDF signals for the ring-opened products following $\alpha$ (\textbf{C0$^{T}$} product) or $\beta$-CC (\textbf{C0} product) cleavage.  Especially between 2 and 3 \AA, the two signals show significant differences which means that they will be clearly distinguishable with UED experiments. In general, all of the observed photoproducts give rise to $\Delta$PDF signals that show substantial differences that can be distinguished experimentally.
Generally, the signals of the photoproducts from the CASSCF dynamics look similar to the signals of the same photoproducts from the TDDFT dynamics. The most important photoproduct, the ring-open structure following $\alpha$-CC cleavage \textbf{C0$^{T}$}, shows a depletion at 1.5 \AA~and 2.3 \AA~and an increase between around 3 and 4 \AA. However, the CASSCF signal shows the same intensity of the two depletion peaks while the TDDFT signals predict a much stronger depletion around 1.5 \AA. 

\begin{figure}[h]
  \begin{center}
    \includegraphics[width=\linewidth]{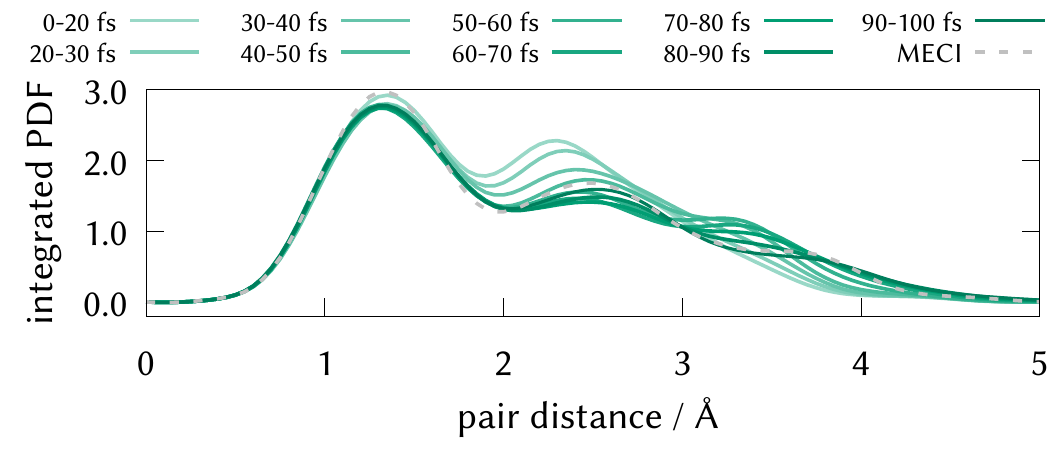}
    \end{center}     
    \caption{%
     Integrated difference pair distribution functions of the \textbf{C0$^T$} photoproduct of cyclobutanone calculated with SA(6)-CASSCF(8,11)/aug-cc-pVDZ.
      }
    \label{fig:greenprod}
\end{figure}

Looking at the integrated signals of the photoproducts, the signal increase of the CASSCF dynamics between 3 and 4 \AA~could be related to the later dissociation of the ring-opened structure, since both dissociated products show a broad increase of intensity around 4 \AA. However, as seen in the populations in Fig.~\ref{fig:photoproducts}, these are only minorly observed within the 100 fs simulation time.
Therefore, we look at the signal of the \textbf{C0$^{T}$} photoproduct in more detail. 
In Fig.~\ref{fig:greenprod}, we report the integrated PDF signal of this structure, integrated over smaller intervals (10 fs). 
It can be observed that the intensity of the peak around 2.3 \AA~decreases over time, stabilising after 50 fs. At the same time, the shoulder between 3 and 4 \AA~gets more pronounced at later times and also moves to higher values, being close to 4 \AA~after 100 fs. 

Comparing it with the signal arising from the optimized MECI structure (Coordinates in Table S1 and representation in Figure S4 in the ESI$^\dagger$), we see a close correspondence with the signals at later times. This is interesting, as the MECI structure is characterized by a linearization of the ring-open chain. 
Therefore, we can associate the shift of the peak towards 4 \AA~after 50 fs during the dynamics, to a linearization of the carbon chain. This increases the O-C$_1$ distance and therefore gives rise to this peak at larger distances. 

\section{Conclusions\label{sec:conclusions}}
The photoproducts formed after the excitation of cyclobutanone to the lower lying n-3s Rydberg state are highly dependant on the quantum chemistry method used to describe the on-the-fly potential energy surfaces. The lifetime of cyclobutanone ranges from 0.5 ps in TD-DFT to 50 fs in CASSCF. Since we do not monitor the electronic state character during the dynamics, we cannot obtain accurate time constants for the n-3s to n$\pi^*$ transition. However, the timescales obtained in this work seem to be anyhow shorter than the 700-900 fs reported by Kuhlman.\cite{Kuhlman2012-exp,symmetry_kuhlman_2012} When using TD-B3LYP-D3 on 48 electronic states including both the singlet and the triplet manifold, cyclobutanone experiences an $\alpha$-cleavage that populates equally the \textbf{C0$^T$} and the \textbf{C2} products, leading about 20\% of the population in each channel after 250 fs. When using the more expensive and more accurate CASSCF method there is a full conversion from cyclobutanone to the $\alpha$-cleavage \textbf{C0$^T$} photoproduct in 100 fs, with a 10\% population going through the \textbf{C3} channel, making the deactivation on CASSCF surfaces about ten times faster than on TD-B3LYP-D3 ones. 

Surprisingly, allowing hops to the ground state does not open new deactivation pathways in TD-B3LYP-D3 and we still do not see the \textbf{C3} product. Therefore, an accurate description of the Rydberg states fundamentally changes the life times as well as the outcome of the photoproduct formation. The TSH/TD-B3LYP behaviour was already predicted by Diau et al. ~\cite{diau2001femtochemistry} to occur at excitation wavelengths below 320 nm, where the predominant channel is \textbf{C2}.

In CASSCF the S$_0$ population is three times higher than in TD-B3LYP-D3 after 100 fs. We would be more inclined to trust the CASSCF results, however the bad statistics indicates that the active space chosen is lacking flexibility to describe every kind of photoproduct formed. 
Due to computational limitations, longer simulation times were not affordable. For the sake of the study of the initial photoinduced ring-opening and photoproduct formation of cyclobutanone, the obtained simulation times are appropriate. In particular with CASSCF, we were able to observe the ring-opening process for the entire ensemble of trajectories. The comparison of the electronic structure methods could, thus, be well carried out on the presented data.

UED spectra show a qualitative picture of the nuclear dynamics. The differences between the electronic structure methods are clearly visible, as expected from the differences in the photoproduct distribution. As the main photoproducts show large differences in the $\Delta$PDF signals, it can be expected that UED experiments will be a successful tool to distinguish the structures and follow the photoproduct formation in real time. 
The time-resolved PDF signals allow us to extract a proposed mechanism from the CASSCF dynamics: $\alpha$-CC cleavage of the molecule is initially driven by the C$_2$ carbon, followed by motion of the carbonyl group and a linearization of the molecule after 50 fs. We can conclude that UED is a very powerful technique, able to probe molecular geometries on the excited state and we are looking forward to see the experimental results on cyclobutanone that will soon be available.

This article is part of a collection aimed at providing theoretically computed electron diffraction spectra to assist spectroscopists in interpreting data gathered on cyclobutanone. The non-adiabatic dynamics community faced a challenge to predict the outcome of an UED experiment within a six-month timeframe. Absent the time constraints set forth by the challenge, had sufficient time been available, we would have improved our TD-DFT non-adiabatic dynamics by employing the more correct SF-TDDFT method, although considering a reduced state manifold to avoid spin contamination problems. Additionally, we would have extended our simulations to longer timescales, conducted assessments on larger active spaces and tested the CAS-CI method to improve the stability of our CASSCF simulations, and ultimately employed the quantum trajectory method DD-vMCG to investigate the influence of quantum effects on the photodissociation of cyclobutanone.
We believe that the main limitation of the performed simulations is with respect to the predicted timescales. Using higher accuracy electronic structure and nuclear dynamics methods can be expected to yield timescales for the deactivation and photoproduct formation in better agreement with the experimental results. However, the quantitative, preliminary photoproduct formation, i.e. $\alpha$C-C bond breaking predicted by CASSCF is in line with previous experimental and computational results.

\section{Supplementary Material}
The input data to reproduce the calculations are available in the form of a zip file. A pdf file is also linked, which contains optimised geometries and frequencies, active space for the CASPT2 energies and CASSCF trajectories, natural transition orbitals, a benchmark on the excited state energies, a reaction path comparing CASSCF and CASPT2, a photodissociation scheme to decide the photoproducts formed, the convergence of the electronic population with the number of trajectories for TSH/CASSCF and calculated electronic scattering signals.

\begin{acknowledgments}
SG acknowledges the Mar\'ia Zambrano grant
for the attraction of international talent (NextGenEU funds) and the USAL grant "Programa Propio C1". 
The authors thank the funding by Spanish Ministry of Science and Innovation (MCIN/AEI/10.13039/501100011033) grant No. PID2020-113147GA-I00. 
This research has made use of the high performance computing resources of the Castilla y León Supercomputing Center (SCAYLE), financed by the European Regional Development Fund (ERDF).
LMI acknowledges the support of the ANR Q-DeLight project (Grant No. ANR-20-CE29-0014) of the French Agence Nationale de la Recherche. All authors thank the COST Actions CA18212 (MD-GAS) and CA21101 (COSY) for fruitful discussions.
\end{acknowledgments}

\section*{Data Availability Statement}

The data that support the findings of this study are available within the article and its supplementary material$^\dagger$.

\section*{References}

\bibliography{mybib} 

\end{document}